\documentclass[twocolumn,showpacs,preprintnumbers,amsmath,amssymb,superscriptaddress]{revtex4}

\usepackage{graphicx}
\usepackage{dcolumn}
\usepackage{bm}
\usepackage{enumerate}

\begin{document}


\title{Metabolic networks are almost nonfractal: A comprehensive evaluation}

\author{Kazuhiro Takemoto}
\email{takemoto@bio.kyutech.ac.jp}
\affiliation{
Department of Bioscience and Bioinformatics, Kyushu Institute of Technology, Iizuka, Fukuoka 820-8502, Japan
}

\date{\today}

\begin{abstract}
Network self-similarity or fractality are widely accepted as an important topological property of metabolic networks; however, recent studies cast doubt on the reality of self-similarity in the networks.
Therefore, we perform a comprehensive evaluation of metabolic network fractality using a box-covering method with an earlier version and the latest version of metabolic networks, and demonstrate that the latest metabolic networks are almost self-dissimilar, while the earlier ones are fractal, as reported in a number of previous studies.
This result may be because the networks were randomized because of an increase in network density due to database updates, suggesting that the previously observed network fractality was due to a lack of available data on metabolic reactions.
This finding may not entirely discount the importance of self-similarity of metabolic networks.
Rather, it highlights the need for a more suitable definition of network fractality and a more careful examination of self-similarity of metabolic networks.
\end{abstract}

\maketitle

\section{Introduction}
{\it Metabolism} can be defined as a series of biochemical reactions, and it is often represented as a network.
Generically, when metabolic networks are depicted in a schematic fashion, metabolites are portrayed as nodes and reactions as edges \cite{Jeong2000,Ma2003,Arita2004}.
In recent years, several new technologies and high-throughput methods have generated a massive quantity of genomic and metabolic network data; thus, the overall picture of the metabolic world has gradually become clearer (reviewed in \cite{Barabasi2004,Albert2005,Takemoto2012a,Winterbach2013}, for example).

In particular, several studies, until now, have demonstrated a scale-free property (i.e., a power-law distribution of the number of links per node or degree) in metabolic networks as well as in other types of networks such as social networks and World Wide Web \cite{Albert2002}.

The scale-free property (power-law degree distribution) is often used in the sense of scale invariance (i.e., self-similarity).
In addition to this, a hierarchical organization (i.e., self-similar nesting of different modules) in complex networks \cite{Ravasz2003a}, including metabolic networks \cite{Ravasz2003b}, also implies network self-similarity.

Song et al. \cite{Song2005} have shown that a number of real-world networks, including metabolic networks, have self-similarity (fractality) using a box-counting method for complex networks, in which a box is defined as a set of nodes between which distances (shortest path lengths) are less than a box size ($l_B$).
In particular, a fractal network shows a power-law relationship in the minimum
number of boxes ($N_B(l_B)$) covering nodes with linear size $l_B$:
\[
N_B(l_B)\propto l_B^{-d_B},
\]
where $d_B$ is a fractal dimension.

According to this previous study, in this paper, network {\it self-similarity} is defined by $N_B(l_B)$ (see Sec. \ref{sec:fractality} for details), and it is synonymous with network {\it fractality}.
However, note that the scale-free property (power-law degree distribution) does not necessarily correspond to network fractality.
In particular, Song et al. \cite{Song2005} have demonstrated that scale-free random networks (i.e., Barab\'asi--Albert (BA) model \cite{Albert2002}) show no fractality while they have power-law degree distributions; in particular, $N_B(l_B)$ of BA model networks exponentially decays with $l_B$. 

Network fractality originates from a negative degree--degree correlation (i.e., disassortativity) \cite{Yook2005} or the repulsion between hubs \cite{Song2006}, and it helps not only to detect functional modules in biological networks but also to understand network evolution \cite{Song2006} and the utility and the redundancy in networked systems \cite{Goh2006}.
Therefore, a number of investigators have accepted self-similarity of metabolic networks, and they have used metabolic networks as an example of fractal networks \cite{Gallos2007a,Gallos2007b,Goh2006,Yook2005,Song2006,Song2005,Sun2014}.

However, network self-similarity has sometimes been criticized \cite{Arita2005}.
Metabolic networks are suggested to be self-dissimilar and scale-rich \cite{Tanaka2004,Tanaka2005} because they are compartmentalized \cite{Takemoto2012b} and the degree distributions among different biosynthetic modules are very different, although the definition of modules (boxes) is different between these studies and previous studies (e.g., \cite{Song2005,Song2006}).

This inconsistent conclusion emphasizes the need for a more careful examination of network fractality.
In particular, previous studies (e.g., \cite{Song2005,Song2006}) have the following problems: consideration of only a few representative organisms and the use of an old dataset of metabolic networks published in 2000 \cite{Jeong2000}.
Thus, metabolic network fractality is still debatable.
In particular, universality of metabolic network fractality is still poorly evaluated.
Moreover, metabolic network analysis strongly depends on the accuracy of metabolic information.
An opposite conclusion may be derived from a comparison between an earlier version and the latest version of metabolic networks \cite{Takemoto2013a}.

The latest data regarding metabolic information are collected in several databases such as the Kyoto Encyclopedia of Genes and Genomes (KEGG) \cite{Kanehisa2014} and the Encyclopedia of Metabolic Pathways (MetaCyc) \cite{Caspi2012}; these databases are widely used and contain the metabolic pathways of many living organisms.

In this study, therefore, we perform a comprehensive evaluation of metabolic network fractality using a box-covering method with an earlier version and the latest version of metabolic networks of a large number of organisms, and we demonstrate that the latest version of metabolic networks show self-dissimilarity while the earlier version of metabolic networks are fractal, as reported in \cite{Song2005,Song2006}.
Furthermore, we discuss a possible origin of the vanishment of network fractality due to database updates, importance of a careful examination of biological networks, more suitable definition of network fractality.


\section{Materials and methods}
\label{sec:method}
\subsection{Selection of organisms}
\label{app:organism}
We used previously published lists of prokaryotes (i.e., archaea and bacteria) \cite{Takemoto2012b} and eukaryotes \cite{Takemoto2013b}.
The metabolic networks in these datasets were well identified and available from the KEGG database \cite{Kanehisa2014}.
To prevent redundancies, when a species had different strains, we proceeded to use the strain whose genome was reported first as the representative strain for that species.
Finally, 172 organisms, including 45 archaea, 60 bacteria, and 67 eukaryotes, were investigated (see Supplemental Material).

\subsection{Construction of metabolic networks}
\label{sec:construction}
The construction of metabolic networks follows similar protocols as those described in a previous study \cite{Takemoto2013a}.

We downloaded XML files (version 0.7.1) containing the metabolic network data of 172 organisms on March 17, 2014 from the KEGG database \cite{Kanehisa2014}.
On the basis of the metabolic information, substrate--product relationships were identified as carbon traces using the latest version \cite{Stelzer2011} and an earlier version \cite{Ma2003} of the metabolic reaction database.
For comparison to previous studies (e.g., \cite{Song2005,Song2006}), the metabolic networks are represented by undirected networks (i.e., substrate graphs) in which the nodes and edges correspond to metabolites and reactions, respectively (i.e., substrate--product relationships are based on atomic mapping \cite{Arita2004}).
Currency metabolites such as H$_2$O, ATP, and NADH were excluded as described previously \cite{Takemoto2007}.
Moreover, the largest (weakly) connected component was extracted from each metabolic network to obtain more accurate calculations of $N_B(l_B)$ (i.e., to avoid bias from isolated components).

The constructed metabolic networks are available as a Supplemental Material.

\subsection{Determination of network fractality}
\label{sec:fractality}
Network fractality is determined by $N_B(l_B)$ \cite{Song2005,Song2006,Song2007}, calculated using a box-covering method.
Covering a network with the minimum possible number of boxes is related to graph coloring \cite{Song2007}, an NP-hard problem; thus, it requires approximation algorithms.
Although Song et al. \cite{Song2007} have proposed several algorithms, we used the compact-box-burning (CBB) algorithm in this study because of the simple implementation of the CBB algorithm.
$N_B(l_B)$ was averaged over 100 realizations.
The selection of box covering algorithms poses little problem because all the algorithm in \cite{Song2007} can find the optimal solution with the similar accuracy.

The previous studies \cite{Song2005,Song2006,Song2007} have demonstrated that fractal and nonfractal networks show $N_B(l_B)\propto l_B^{-d_B}$ and $N_B(l_B)\propto \exp(-l_B)$ (because of a small-world property or logarithmic increase of the average diameter of a network with the total number of nodes $N$), respectively.
We discriminate network fractality by comparing between the coefficients of determination, a measure of goodness-of-fit, of a power-law fit ($R^2_{\mathrm{pow}}$) and exponential fit ($R^2_{\mathrm{exp}}$) between $N_B(l_B)$ and $l_B$.
$R^2$ is not influenced by sample size because it is an effect size in statistics.

$R^2_{\mathrm{pow}}$ and $R^2_{\mathrm{exp}}$ were calculated from the {\it lm} function in the R software (www.r-project.org) using the regression formulas $\ln N_B(l_B)=-d_B\ln l_B + b_1$ and $\ln N_B(l_B)=a_2 l_B + b_2$, respectively.

$R^2_{\mathrm{pow}} > R^2_{\mathrm{exp}}$ and $R^2_{\mathrm{pow}} < R^2_{\mathrm{exp}}$ indicate a fractal network and nonfractal one, respectively.

\section{Results}

\subsection{Metabolic networks are almost self-dissimilar}
\label{sec:nonfractal}
\begin{figure}[tbhp]
\begin{center}
	\includegraphics{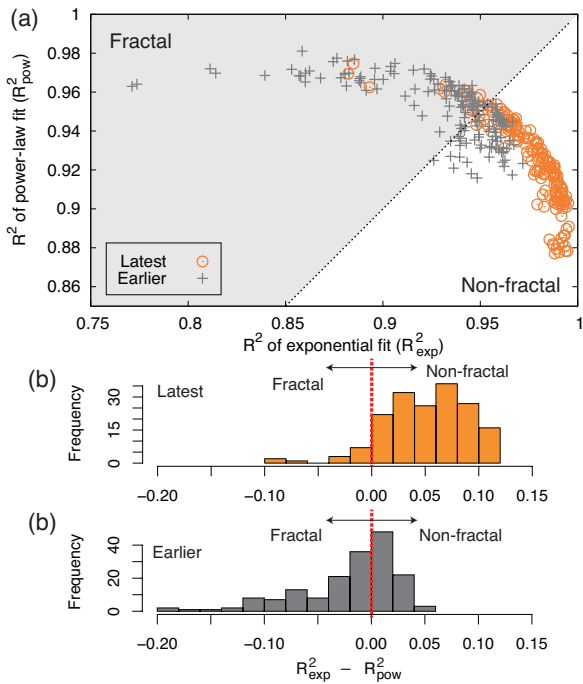}  
	\caption{(Color
 online) Comparison between the coefficients of determination of an exponential fit ($R_{\mathrm{exp}}^2$) and power-law fit ($R_{\mathrm{pow}}^2$).
 (a) Scatter plot of $R_{\mathrm{exp}}^2$ versus $R_{\mathrm{pow}}^2$.
 Distributions of $R_{\mathrm{exp}}^2 - R_{\mathrm{pow}}^2$ in the earlier version (b) of latest version (c) of metabolic networks.
 }
	\label{fig:compare_exp_pow_compact}
\end{center}
\end{figure}

\begin{figure*}[tbhp]
	\includegraphics{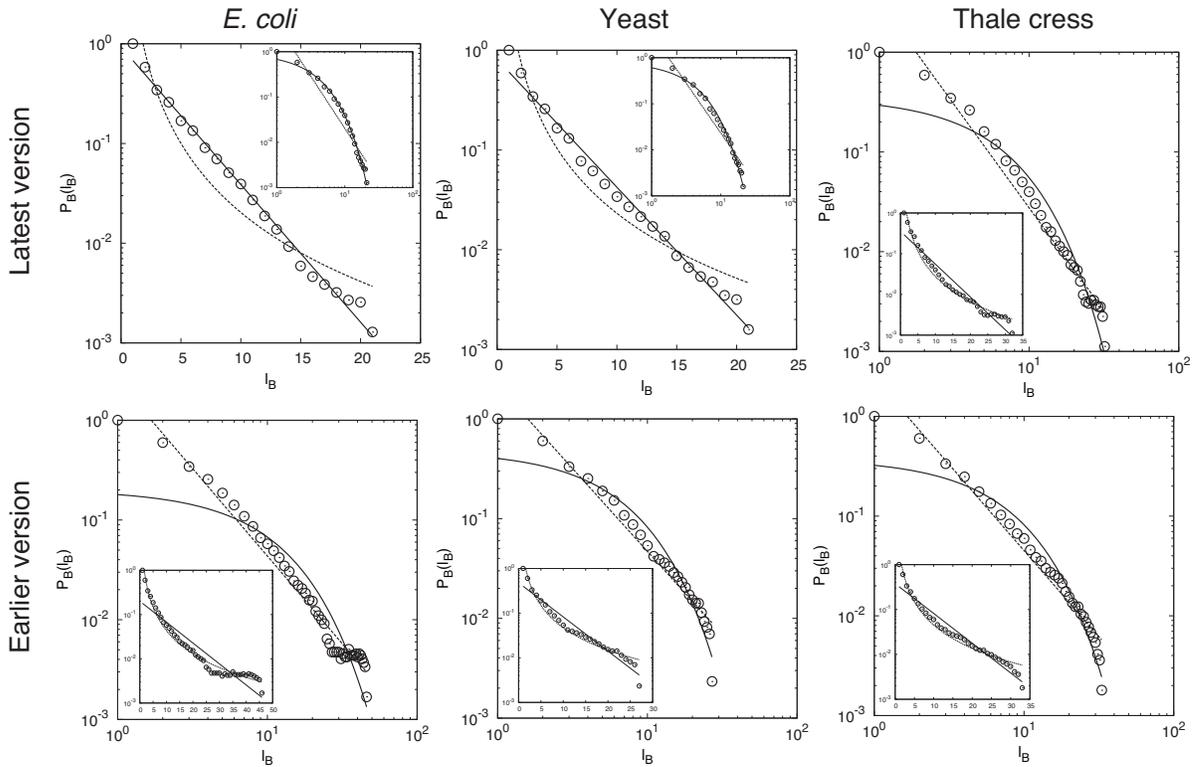}
	\caption{Scatter plots of the normalized number of boxes (i.e., $P_B(l_B)=N_B(l_B)/N$) versus box size $l_B$ in three representative organisms, obtained from the latest version (upper row) and earlier version (lower row) of metabolic networks: {\it Escherichia coli} (left column), {\it Saccharomyces cerevisiae} (yeast) (middle column), {\it Arabidopsis thaliana} (thale cress) (right column).
	Solid and dashed lines indicate exponential fits and power-law fits, respectively.}
	\label{fig:box_distribution}
\end{figure*}

A comparison between $R^2_{\mathrm{pow}}$ and $R^2_{\mathrm{exp}}$ (Fig. \ref{fig:compare_exp_pow_compact} (a)) demonstrates that the latest version of metabolic networks are not fractal because $R^2_{\mathrm{pow}}<R^2_{\mathrm{exp}}$ ($p=2.2\times10^{-16}$ using the one-sample $t$-test) (see also Fig. \ref{fig:compare_exp_pow_compact} (b)).
In particular, $N_B(l_B)$ is described by an exponential function rather than a power-law function (mean $R^2_{\mathrm{exp}}$ is 0.97) (Fig. \ref{fig:box_distribution}).
As a representative example, the cases of {\it Escherichia coli}, yeast, and thale cress are shown.
However, several exceptions were observed.
For example, the latest metabolic network of thale cress, a plant, show the self-similarity.

The network representation in this study is slightly different from that in previous studies (e.g., \cite{Song2005,Song2006}).
We defined edges as substrate--product pairs on the basis of carbon trances (see Sec \ref{sec:construction}) in this study.
This approach was inspired by Arita's study \cite{Arita2004}, in which he pointed out that the pathways computed in the classical manners (i.e., network representations without consideration of atomic traces) do not conserve their structural moieties and, therefore, do not correspond to biochemical pathways on the traditional metabolic map.
In addition to this, it remains possible that the definition of currency metabolites, such as water and ATP, is slightly different from that in previous studies (e.g., \cite{Song2005,Yook2005,Song2006,Goh2006,Gallos2007b,Sun2014}).
In particular, we considered the currency metabolites in our previous study \cite{Takemoto2007}, which are defined according to famous previous studies on metabolic networks (e.g., \cite{Jeong2000,Ma2003,Arita2004}).
Since the previous studies used the metabolic network data in Ref. \cite{Jeong2000}, it is expected that the definition is almost similar between our study and the previous studies.
However, we could not conclude whether the definition is really similar because the definition has not been clearly described in the previous studies on network factuality.

To show that differences in network representation pose few problems, we also evaluated network factuality of the earlier version of the metabolic networks, using the box-counting method \cite{Song2007}.
The earlier version of metabolic networks are fractal because $R^2_{\mathrm{pow}}>R^2_{\mathrm{exp}}$ ($p=1.1\times10^{-7}$ using the one-sample $t$-test) (see also Fig. \ref{fig:compare_exp_pow_compact} (c)), as reported in a number of previous studies (e.g., \cite{Song2005,Song2006}).
Especially, $N_B(l_B)$ can be accurately described using a power-law function (mean $R^2_{\mathrm{pow}}$ is 0.95) (Fig. \ref{fig:box_distribution}).
This result is consistent with the finding reported in the previous studies \cite{Song2005,Song2006}; thus, it indicates that the procedures used for data analysis (e.g., the definition of currency metabolites) in this study were not problematic.
On the other hand, the difference in analysis results between the earlier version and latest version of metabolic networks implies that the observed network fractality might result from a lack of data on metabolic reactions.

\subsection{Differences in network measures between the earlier and latest versions of metabolic networks}
\label{sec:diff}
We here discuss the effect of updates of the metabolic reaction database on network fractality.

Inspired by a relationship between disassortativity (i.e., negative degree--degree correlation) and network self-similarity \cite{Yook2005}, we first focused on the difference of an assortative coefficient $r$ \cite{Newman2002}, defined as Pearson product-moment correlation coefficient of degree--degree correlations, between the earlier version ($r_{\mathrm{earlier}}$) and latest version ($r_{\mathrm{latest}}$) of metabolic networks: $\Delta r = r_{\mathrm{latest}}-r_{\mathrm{earlier}}$.
However, such a difference could not be concluded ($p=0.41$ using the one-sample $t$-test) although both versions of metabolic networks show a weak disassortativity (mean $r$ of the earlier version and latest version of the networks are $-0.011$ ($p=0.0082$ using the one-sample $t$-test) and $-0.0080$ ($p=0.0054$ using the one-sample $t$-test), respectively).
This result implies that the vanishment of network fractality cannot be explained in terms of assortative mixing in networks.

The database updates resulted in the metabolic networks of the latest version being larger than those of the earlier version.
Thus, we next investigated a simpler network property: network density $D$, defined as $E/N$, where $E$ and $N$ are the number of nodes and the number of links, respectively.

We here considered the change ratio of a network measure $X$, defined as $R_X= X_{\mathrm{latest}}/X_{\mathrm{earlier}}-1$, where $X_{\mathrm{latest}}$ and $X_{\mathrm{earlier}}$ are $X$ obtained from the latest and earlier versions of metabolic networks, respectively.
This definition had a limitation in that it did not consider the loss of nodes and edges because of the database update; however, this discrepancy was minor.
The latest metabolic network included approximately 98\% and 95\% of nodes and edges, respectively, contained in the earlier version \cite{Takemoto2013a}.

$D_{\mathrm{latest}}$ is significantly larger than $D_{\mathrm{earlier}}$ ($p<2.2\times 10^{-16}$ using the one-sample $t$-test).
Furthermore, we found a negative correlation between $R_D$ and $\Delta_{R^{2}_{\mathrm{pow}}}$, defined as $[R^{2}_{\mathrm{pow}}]_\mathrm{latest}-[R^{2}_{\mathrm{pow}}]_\mathrm{earlier}$, where $[R^{2}_{\mathrm{pow}}]_\mathrm{latest}$ and $[R^{2}_{\mathrm{pow}}]_\mathrm{earlier}$ are $R^{2}_{\mathrm{pow}}$ obtained from the latest version and earlier version of metabolic networks, respectively.
This result suggests that the increase in network density (i.e., database updates) decreases the self-similarity in metabolic networks.

\begin{figure}[tbhp]
	\includegraphics{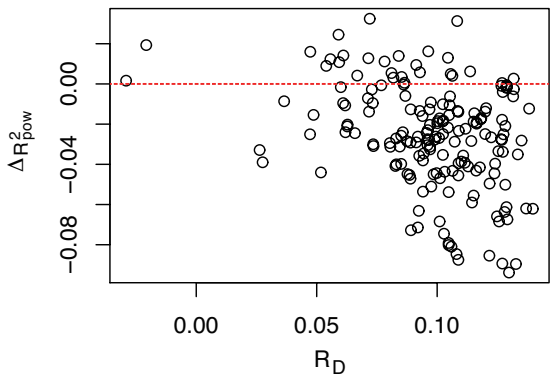}
	\caption{(Color online) Negative correlation between $\Delta_{R^{2}_{\mathrm{pow}}}$ and $R_D$ (Spearman's rank correlation coefficient $r_s=-0.30$ and $p=5.2\times 10^{-5}$).}
	\label{fig:diff_network_param}
\end{figure}

Taken together, we speculate that the decrease in network fractality (i.e., $R^{2}_{\mathrm{pow}}$) due to updating of the database occurred for the following reasons.
In general, network fractality decreased slightly because metabolic networks may be randomized due to the addition of links among previously existing nodes.
In this case, a number of structural properties are changed.
In particular, Song et al. \cite{Song2006} have pointed out the importance of diameter in the origin of network factuality.
Thus, an addition of links among boxes, detected in the earlier version of metabolic networks, decreases the network diameter and average shortest path length; as a result, it collapses a fractal structure (i.e., it leads to an exponential decay of $N_B(l_B)$).
This speculation may be supported by the observed positive correlations of $\Delta_{R^{2}_{\mathrm{pow}}}$ with the absolute change ratio of the average shortest path length $L$ (i.e., $|R_L|$) (Spearman's rank correlation coefficient $r_s=-0.36$ and $p=1.5\times 10^{-6}$) and absolute change ratio of the diameter $d$ (i.e., $|R_d|$) ($r_s=-0.68$ and $p<2.2\times 10^{-16}$) (see Supplemental Material).
Note that an absolute change ratio was used in this case because of these network measures have negative change ratios. 

\subsection{A part of eukaryotic metabolic networks are fractal, but it is still debatable}
Although the latest version of metabolic networks tends to be nonfractal, there are several exceptions (Figs. \ref{fig:compare_exp_pow_compact} and \ref{fig:box_distribution}).
To reveal a tendency of these exceptions, we investigated metabolic network fractality in grater details, according to species classifications.

\begin{figure}[tbhp]
	\includegraphics{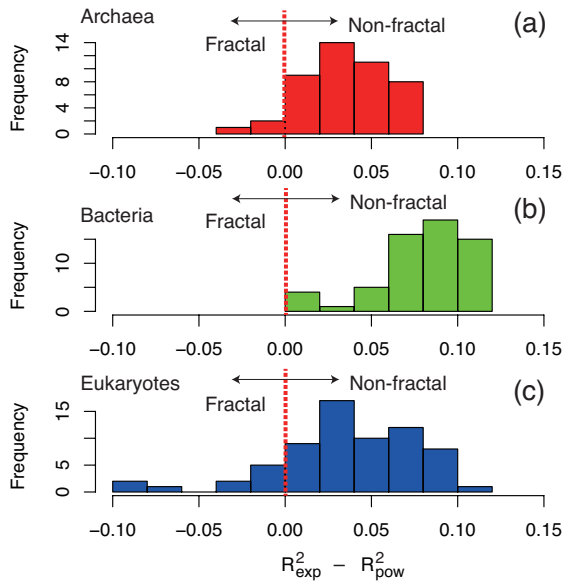}
	\caption{(Color
 online) Distributions of $R_{\mathrm{exp}}^2 - R_{\mathrm{pow}}^2$, obtained from the latest version of metabolic networks, in archaea (a), bacteria (b), and eukaryotes (c).}
	\label{fig:hist_domain}
\end{figure}

We first focused on three domains of life: archaea, bacteria, and eukaryotes (Fig. \ref{fig:hist_domain}).
The archaeal and bacterial (i.e., prokaryotic) metabolic networks are significantly self-dissimilar because $R_{\mathrm{exp}}^2 > R_{\mathrm{pow}}^2$ in both cases of archaea and bacteria ($p=1.6\times 10^{-11}$ and $p<2.2\times 10^{-16}$ using the one-sample $t$-test, respectively).
Overall, the eukaryotic metabolic networks are also nonfractal because $R_{\mathrm{exp}}^2 > R_{\mathrm{pow}}^2$ ($p=2.2\times 10^{-9}$); however, the distribution of $R_{\mathrm{exp}}^2 - R_{\mathrm{pow}}^2$ in the eukaryotic networks is broader than that in the prokaryotic ones and more exceptions are observed in the eukaryotic case.
A degree of metabolic network change due to database updates may be not able to explain this difference in the distributions because the mean change ratio $R_D$ in prokaryotes (archaea and bacteria) (0.099) is almost equivalent to that in eukaryotes (0.096) ($p=0.52$ using the two-sample $t$-test).
That is, it is difficult to conclude that eukaryotic metabolic networks are relatively self-similar because they are relatively unchanged despite database updates.

\begin{figure}[tbhp]
	\includegraphics{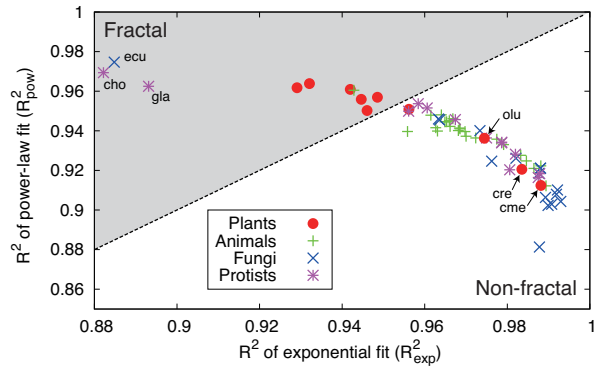}
	\caption{(Color online) Scatter plot of $R_{\mathrm{exp}}^2$ versus $R_{\mathrm{pow}}^2$ in the latest version of eukaryotic metabolic networks.}
	\label{fig:compare_exp_pow_compact_eukaryotes}
\end{figure}

Thus, we next investigated on eukaryotic metabolic network fractality in grater details, according to a eukaryotic classification (Fig. \ref{fig:compare_exp_pow_compact_eukaryotes}), defined in the KEGG database (i.e., plants, animals, fungi, and protists), and found that plant metabolic networks show fractality although the following plant species are exceptions: {\it Ostreococcus lucimarinus} (olu), {\it Chlamydomonas reinhardtii} (cre), and {\it Cyanidioschyzon merolae} (cme).
Since these plant species are microorganisms, they can be distinguished from the other plants, which are higher plants such as trees and flowers.
Thus, it concluded that metabolic networks of higher plants are fractal because $R_{\mathrm{pow}}^2 > R_{\mathrm{exp}}^2$ ($p=0.034$ using the one-sample $t$-test).
However, a degree of metabolic network change due to database updates (i.e., mean $R_D$) is different between the higher plants (0.081) and the other eukaryotes (0.098) ($p=0.010$ using the two-sample $t$-test); thus, it remains possible that metabolic networks of higher plants are still self-similar because they show less change despite database updates.

On the other hand, metabolic networks are self-dissimilar (i.e., $R_{\mathrm{exp}}^2 > R_{\mathrm{pow}}^2$) in animals and fungi ($p=1.2\times 10^{-7}$ and $p=4.8\times 10^{-6}$ using the one-sample $t$-test, respectively); however, we cannot determine whether protist metabolic networks are fractal or nonfractal because $R_{\mathrm{pow}}^2 = R_{\mathrm{exp}}^2$ ($p=0.55$ using the one-sample $t$-test).
In particular, the metabolic networks of {\it Encephalitozoon cuniculi} (ecu), {\it Cryptosporidium hominis} (cho), and {\it Giardia lamblia} (gla) show the self-similarity, exceptionally.
However, the metabolic networks of these eukaryotes were almost unchanged due to database updates.
Especially, mean $R_D$ of these eukaryotes is 0.0057, and it is significantly smaller than that of other eukaryotes (0.098).

Taken together, the metabolic networks of eukaryotes are also self-dissimilar.
Although metabolic network fractality can be concluded in a part of eukaryotes (higher plants, in particular); however, it is still debatable.

\section{Discussion}
In this study, we showed that metabolic networks are almost self-dissimilar, contrary to a number of previous studies \cite{Gallos2007a,Gallos2007b,Goh2006,Yook2005,Song2006,Song2005,Sun2014}.
Rather, this result strongly supports the scale-richness and self-dissimilarity in metabolic networks, proposed by Tanaka \cite{Tanaka2004,Tanaka2005}.
The definition of modules or boxes in Tanaka's studies \cite{Tanaka2004,Tanaka2005} is arbitrary (although it is based on biological knowledge), and it is different from that in previous studies (e.g., \cite{Song2006,Song2005}), which used box-covering methods.
Thus, it is difficult to make an easy comparison between these previous studies.
In addition to this, these previous studies only focused on a limited number of species. 
Our comprehensive evaluation provided a more conceiving evidence of the self-dissimilarity of metabolic networks by avoiding these limitations in these previous studies.

We demonstrated that the previously observed network fractality was probably due to a lack of available data on metabolic reactions.
Especially, metabolic networks have not been fully understood; thus, there is a need for a more careful examination in data analysis in the future. For example, enzyme promiscuity \cite{Khersonsky2010}, which implies that enzymes can catalyze multiple reactions, act on more than one substrate, or exert a range of suppressions \cite{Patrick2007}, in which an enzymatic function is suppressed by overexpressing enzymes showing originally different functions, suggests the existence of many hidden metabolic reactions.
Consideration of these hidden metabolic reactions is important for understanding metabolic network fractality.
In addition to this, these experimental studies suggest that metabolic networks are more flexible than previously thought.
Moreover, previous theoretical and data analytic studies \cite{Takemoto2012b,Minnhagen2008,Bernhardsson2010,Lee2012} argue that specific features in the networks are weakly correlated with a system-specific purpose, function, or causal chain.
These results imply that metabolic networks more randomly constructed than previously thought, and they confirm our finding that an increase of network density enhances network randomization (Fig. \ref{fig:diff_network_param}).

In addition to this, the earlier version of metabolic networks show self-similarity although it shows a very weak disassortativity, and the network fractality vanished due to database updates although the assortative coefficient was unchanged (Sec. \ref{sec:diff}), this result implies a limitation of disassortativity as determination factor of network self-similarity \cite{Yook2005}, and it underscores the need for a more careful discussion of the origin of self-similarity of complex networks.
Especially, it may suggest the existence of different possible origins of network fractality. 

This study does not discount the importance of self-similarity of metabolic networks.
Rather, it emphasizes the need for a more suitable definition of network fractality and a more careful examination of self-similarity of metabolic networks.
For example, this study does not consider several important properties of metabolic networks, as do many other works on metabolic network analyses: reaction stoichiometry, the direction of reaction (i.e., reversible versus irreversible), chemical structure of metabolites, and gene clusters.
It will be important to propose box-covering (renormalization) methods that also consider such kinds of biological information in metabolic networks.
In this context, methods for finding modular architecture (i.e., biologically meaningful boxes) of metabolic pathways based on gene clusters and chemical transformation patterns \cite{Muto2013,Kanehisa2013} may be useful.
Using these definitions, we may be able to evaluate biologically understandable fractality in metabolic networks and other biological networks. 

\section*{Acknowledgments}
This study was supported by a Grant-in-Aid for Young Scientists (A) from the Japan Society for the Promotion of Science (Grant No. 25700030).
K.T. was partly supported by Chinese Academy of Sciences Fellowships for Young International Scientists (Grant No. 2012Y1SB0014), and the International Young Scientists Program of the National Natural Science Foundation of China (Grant No. 11250110508).


\end{document}